\documentclass[12pt]{article}

\usepackage{newtxtext,newtxmath}

\usepackage{graphicx}

\usepackage[letterpaper,margin=1in]{geometry}

\renewenvironment{abstract}
	{\quotation}
	{\endquotation}

\date{}

\makeatletter
\renewcommand{\fnum@figure}{\textbf{Figure \thefigure}}
\renewcommand{\fnum@table}{\textbf{Table \thetable}}
\makeatother

\usepackage{scicite}

\usepackage{url}

\def\scititle{
	Quantifying the safe operating space for the Amazon rainforest under climate change and deforestation
}
\title{\bfseries \boldmath \scititle}

\author{
	Jonathan~Krönke$^{1,2\ast}$,
	Arie~Staal$^{3}$,
	Jonathan~F.~Donges$^{1,4,5}$,\\
    Johan~Rockström$^{1,4,6}$, and
    Nico~Wunderling$^{1,2,7\ast}$\and
	\small$^{1}$Earth Resilience Science Unit, Potsdam Institute for Climate Impact Research, \and 
    \small Member of the Leibniz Association, Potsdam \& 14412, Germany.\and
	\small$^{2}$Center for Critical Computational Studies, Goethe University Frankfurt, \and
    \small Frankfurt am Main \& 60629 Germany\and
    \small$^{3}$Copernicus Institute of Sustainable Development, Utrecht University, Utrecht \& 3508TC, Netherlands\and
    \small$^{4}$Stockholm Resilience Centre, Stockholm University, Stockholm \& SE-106 91, Sweden\and
    \small$^{5}$Max Planck Institute of Geoanthropology, Jena \& 07745, Germany\and
    \small$^{6}$Institute for Earth and Environmental Systems, University of Potsdam, Potsdam \& 14469, Germany\and
    \small$^{7}$Senckenberg Research Institute and Natural History Museum, \and
    \small Member of the Leibniz Association, Frankfurt am Main \& 60325, Germany\and
	\small$^\ast$Corresponding authors. Emails: jonathan.kroenke@pik-potsdam.de, wunderling@c3s.uni-frankfurt.de\and
}

\begin{document} 

\maketitle
\newpage

\begin{abstract} \bfseries \boldmath
The Amazon rainforest is considered one of the core tipping elements in the climate system with a potential tipping point from rainforest to savannah between 2 and 6~$^\circ$C of global warming. However, ongoing deforestation constitutes an additional major threat to the Amazon rainforest that acts simultaneously to undermine the stability of the rainforest. Both effects could synergistically compound and lower the overall threshold in global warming and deforestation when tipping points may be crossed. Here, we quantify the safe operating space of the Amazon rainforest, which we define as the joint global warming and deforestation conditions where resilience of the system as a whole is preserved. Based on the underlying environmental data from a global climate model, we use a reduced complexity model and explicitly take into account the adaptive capacities of the forest as well as the atmospheric moisture recycling. We quantify that under current conditions of around $1.4~^\circ$C of global warming and around 17~\% of deforestation, more than a third of the Amazon rainforest is at high risk of crossing critical thresholds. We therefore conclude that the Amazon rainforest may have already left its safe operating space. Furthermore, we find that the historic and projected deforestation pattern could be particularly detrimental. Our results support the need for ambitious climate action to hold the Paris climate target and also nature protection to end net deforestation.
\end{abstract}\vspace{1cm}

\noindent

The Amazon rainforest is not only the most biodiverse region of the planet holding over 10~\% of Earth's terrestrial biodiversity \cite{flores_critical_2024}, but has also been an important sink for human $\mathrm{CO}_2$ emissions, storing over 100~GtC \cite{armstrongmckay_exceeding_2022, gatti_amazonia_2021, malhi_regional_2006}. Global-warming induced water stress in the Amazon due to the decrease of overall water availability as well as an increase in intensity and frequency of droughts \cite{cook_twentyfirst_2020, duffy_projections_2015} threatens this regulating capacity of the Earth and it has been found to decline regionally \cite{gatti_amazonia_2021}. Further forest loss would significantly reduce the $\mathrm{CO}_2$ uptake capacity of the climate system in addition to a distressing loss of biodiversity. Furthermore, the Amazon rainforest is considered one of the core tipping elements of the climate system \cite{armstrongmckay_exceeding_2022, lenton_tipping_2008} with a potential tipping point at $2-6~^\circ$C and $3.5~^\circ$C as a best estimate \cite{armstrongmckay_exceeding_2022}. Earth System Models with and without dynamic vegetation usually do not show one singular tipping point. Instead they show regional-scale forest dieback with varying spatial extent and in different regions \cite{terpstra_assessment_2025, parry_evidence_2022, drijfhout_catalogue_2015, melnikova_amazon_2025}.\par 
While the Amazon rainforest constitutes a complex ecosystem with numerous feedbacks \cite{flores_feedback_2022}, forest-rainfall interactions are of particular importance and are often discussed in the context of Amazon tipping \cite{zemp_selfamplified_2017, brando_tipping_2025}. Around a third of the rainfall in the Amazon basin is recycled partly by moisture transpired from vegetation, typically travelling 100s to 1000s$~\mathrm{km}$ before raining down again \cite{staal_forestrainfall_2018}. This moisture recycling acts primarily along the dominant wind direction from east to west and is particularly strong during the dry season \cite{boers_deforestationinduced_2017, staal_forestrainfall_2018}. Loss of tree coverage in one region can thus not only significantly reduce the rainfall in remote regions downwind on average, but also intensify the dry season \cite{flores_critical_2024}. Additionally, fire feedbacks can explain observed bimodality of forest and savanna locally ($\sim1~\mathrm{km}$ \cite{staal_hysteresis_2020}) leading to the hypothesis of local (irreversible) tipping initiated by water stress \cite{vannes_tipping_2014, staver_global_2011, hirota_global_2011}. Insufficient consideration of fire and moisture recycling feedbacks could explain why such irreversible tipping is often not observed in dynamic global vegetation models \cite{nian_rainfall_2024}. This justifies the need for complementing approaches based on a combination of known key mechanisms and (semi-)empirical data from observations and Earth System Model output.\par Furthermore, plants are often adapted to their local environmental conditions and it has been suggested that drought impacts will be experienced by plants in relation to the relative departure from their long-term environmental conditions \cite{phillips_drought_2009, tavares_basinwide_2023}. Due to that, the eastern regions of the Amazon with high rainfall seasonality inhabit trees that established deep roots to overcome periods of water stress \cite{fan_hydrologic_2017, sakschewski_variable_2021}.\par
In addition to global-warming induced water stress, deforestation represents a further human-induced pressure. Large-scale deforestation taking place across the Amazon rainforest \cite{staal_synergistic_2015, franco_how_2025, hajdu_deforestation_2025} reduces atmospheric moisture recycling and undermines its overall resilience. The tipping point for the effects of deforestation has been estimated at $20-25~\%$ of deforestation together with 1.5-2.0$~^\circ$C of global warming \cite{lovejoy_amazon_2018}. A possible scenario for the Amazon rainforest would thus be: local precipitation reductions from climate change may exceed adaptation thresholds and initiate local forest dieback. This local forest tipping in combination with deforestation reduces precipitation in downstream areas through diminished atmospheric moisture recycling thereby decreasing their resilience or even initiating cascades and a larger scale tipping of the Amazon rainforest \cite{zemp_selfamplified_2017, wunderling_recurrent_2022}.\par
The interaction of multiple pressures on the Amazon rainforest demands a combined assessment of its resilience. Among many drivers \cite{flores_feedback_2022}, deforestation and precipitation changes due to climate change were identified as major drivers of potential critical transitions in the Amazon rainforest \cite{flores_critical_2024}. A combined assessment under consideration of these two drivers can be viewed as a quantification of the safe operating space of the Amazon rainforest \cite{scheffer_creating_2015} in analogy to the safe operating space for the whole planet, delineated by the planetary boundaries \cite{rockstrom_safe_2009}. The safe operating space of the Amazon rainforest would be the combined climate change and deforestation conditions where the risk of crossing a larger scale or even system-wide tipping point is low or where dangerous levels of forest loss are not yet reached \cite{rockstrom_planetary_2009}, thereby preserving most of the resilience of the Amazon rainforest as a whole. Outside of the safe operating space, resilience losses become detrimental. Determining the boundaries between the safe and dangerous regions is thus key for the limitation of, potentially irreversible, damage to the Amazon rainforest. Following the precautionary principle, individual boundaries of the safe operating space could be set to $2~^\circ$C and to 20~\% deforestation as these correspond to the minima of current tipping point estimation ranges established by \cite{flores_critical_2024}. However, the stress from both drivers occurs simultaneously and interferes via moisture recycling, creating the need for a combined quantification of the safe operating space. Here, we provide such an improved assessment by conducting a safe operating space quantification with explicit consideration of moisture recycling and relate current levels of around $1.4~^\circ$C of global warming \cite{wmo_state_2025} and around 17~\% of forest loss \cite{gatti_amazonia_2021} (mainly by deforestation \cite{flores_critical_2024}) to that safe operating space.\par
\begin{figure} 
	\centering
	\includegraphics[width=0.8\textwidth]{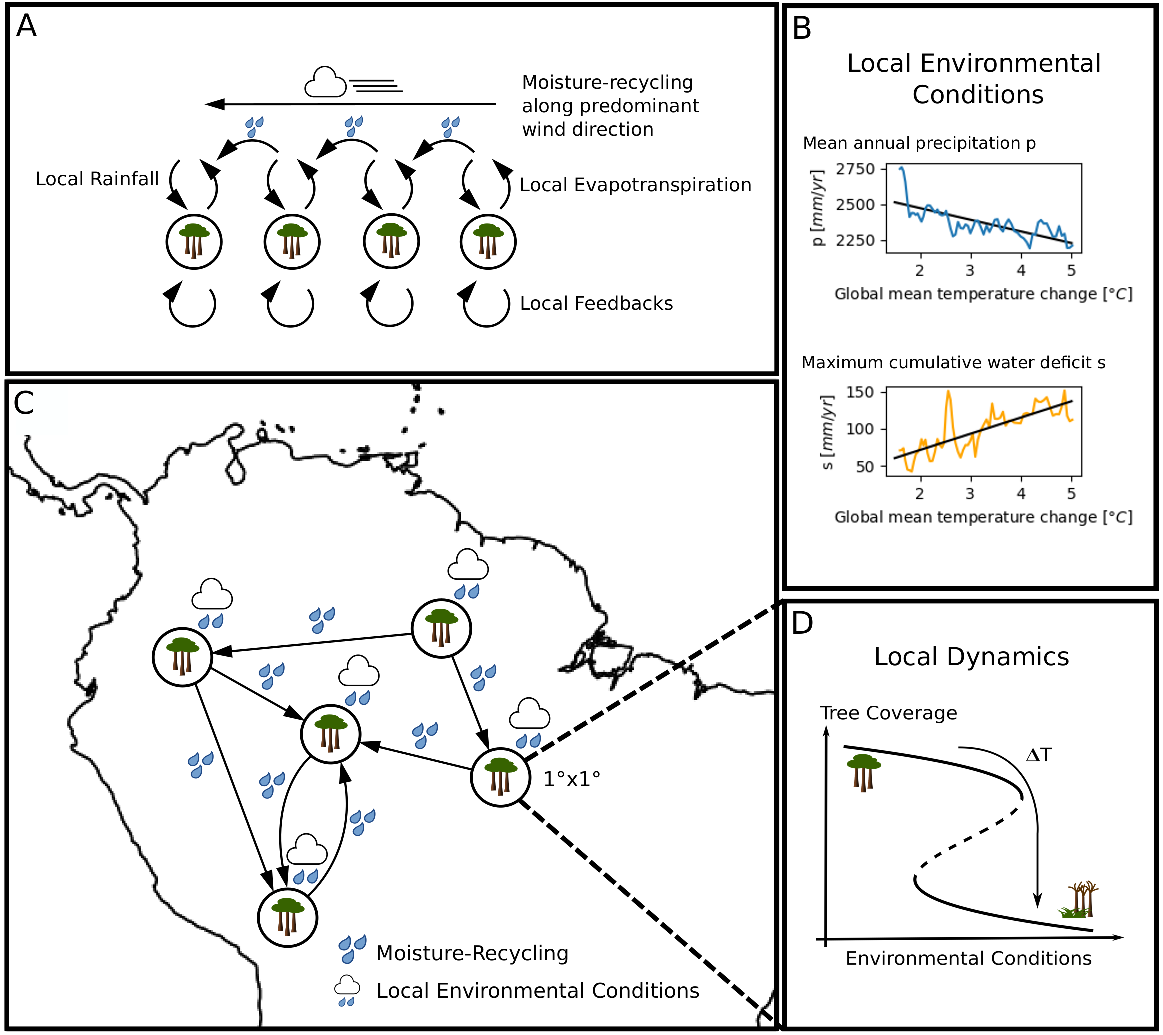} 

	\caption{\textbf{Illustration of the relevant dynamics of the network model.}
	    (\textbf{A}) The relation between local feedbacks and moisture recycling. While local feedbacks are only relevant in one grid cell, moisture recycling along the predominant wind direction is responsible for a large-scale feedback. (\textbf{B}) The local environmental conditions depend on the local mean annual precipitation and maximum cumulative water deficit as well as the moisture recycling received from other cells (see panel c). The individual response of the mean annual precipitation and maximum cumulative water deficit to global warming is obtained from a linear regression to the simulation runs (see methods). (\textbf{C}) Simplified illustration of the dynamical network model with five depicted cells (the full network of all cells includes 416 cells) where the individual cells are connected in a complex network of moisture-flows [see also figure \ref{supfig:networks}]. (\textbf{D}) Bifurcation diagram of the local dynamics in each cell. If the local environmental conditions change sufficiently, local tipping to a degraded state occurs.}
	\label{fig:example} 
\end{figure}

We assess the safe operating space of the Amazon rainforest using a reduced-complexity model of the relevant feedbacks and tipping dynamics with two alternative tree cover states, forest and degraded, on a grid with 416 cells each having a size of around $1^\circ\times1^\circ$ \cite{kronke_dynamics_2020, wunderling_modelling_2021, wunderling_recurrent_2022, wunderling_pinpointing_2025}. The model is parameterized using historical and CMIP6 simulation results from the Earth System Model NorESM2 \cite{seland_overview_2020}. The equilibrium states of an individual forest cell with respect to local environmental conditions are shown in Fig.~\ref{fig:example}D. Note that there is a range of environmental conditions where both states are stable. These local environmental conditions include two relevant features: the mean annual precipitation as a measure for overall water supply and the maximum cumulative water deficit as a combined measure for the intensity and length of the dry season \cite{malhi_exploring_2009}. Recycled moisture from other cells is included in the local environmental conditions. These moisture flows among grid cells are derived using a moisture-tracking algorithm applied to the NorESM projections \cite{staal_global_2025}. The resulting changes in mean annual precipitation and maximum cumulative water deficit are subsequently quantified. Local tipping thresholds are then calibrated based on past variability in those two measures, as simulated in the historical NorESM experiments. Consequently, when environmental conditions deviate increasingly from their long-term states, forest resilience declines until adaptive thesholds are exceeded and tipping occurs. For each cell and moisture recycling link, we derive the response to global warming from the NorESM simulation runs corresponding to three Shared Socioeconomic Pathway (SSP2-4.5, SSP3-7.0, and SSP5-8.5) scenarios (Fig.~\ref{fig:example}B [see also \cite{methods}]). This also enables us to extrapolate to global warming levels that are not reached in the simulations. From the grid-cell level precipitation responses for each scenario we calculate a normal distribution and employ an ensemble of $N=16$ members in each simulation. Deforestation is represented by artificially converting selected grid cells to a low tree-cover state. To assess how directional moisture flows influence system responses, we employ four different spatial patterns of deforestation. First, we include a scenario-based pattern that reflects increased deforestation along road and infrastructure expansions \cite{soares-filho_modelling_2006}. Second and third, we employ two idealized extreme cases - an east-to-west and a west-to-east clearing pattern. And fourth, a randomized deforestation pattern is applied as an additional robustness test.\par
To quantify the safe operating space, we simulate the Amazon for combinations of deforestation between 0 and 50~\% of the forest area and climate change impacts on mean annual precipitation and maximum cumulative water deficit that represent global warming values between $0~^\circ\mathrm{C}$ and $10~^\circ\mathrm{C}$. The initial state of our simulation determines the deforestation $D$ and the final state determines the overall forest loss $L$ in percent of the total forest area across the Amazon. The difference is due to tipping $T$ ($L=D+T$). Note that tipping does not include forest loss from deforestation, e.g. tipping of $T=5~\%$ can still add up to an overall forest loss of $L=20~\%$ in combination with deforestation of $D=15~\%$. This distinction is important to better seperate nonlinear effects of moisture recycling from linear forest losses due to deforestation in quantifying the safe operating space. Crossing of local tipping points can result from a change in the local environmental conditions due to global warming or from cascading transitions initiated by tipping or deforestation in other cells. Further details on the methods can be found in the Supplementary Information.

\subsection*{Individual safe boundaries for climate change and deforestation}
\begin{figure}
  \centering
  \includegraphics[width=1\textwidth]{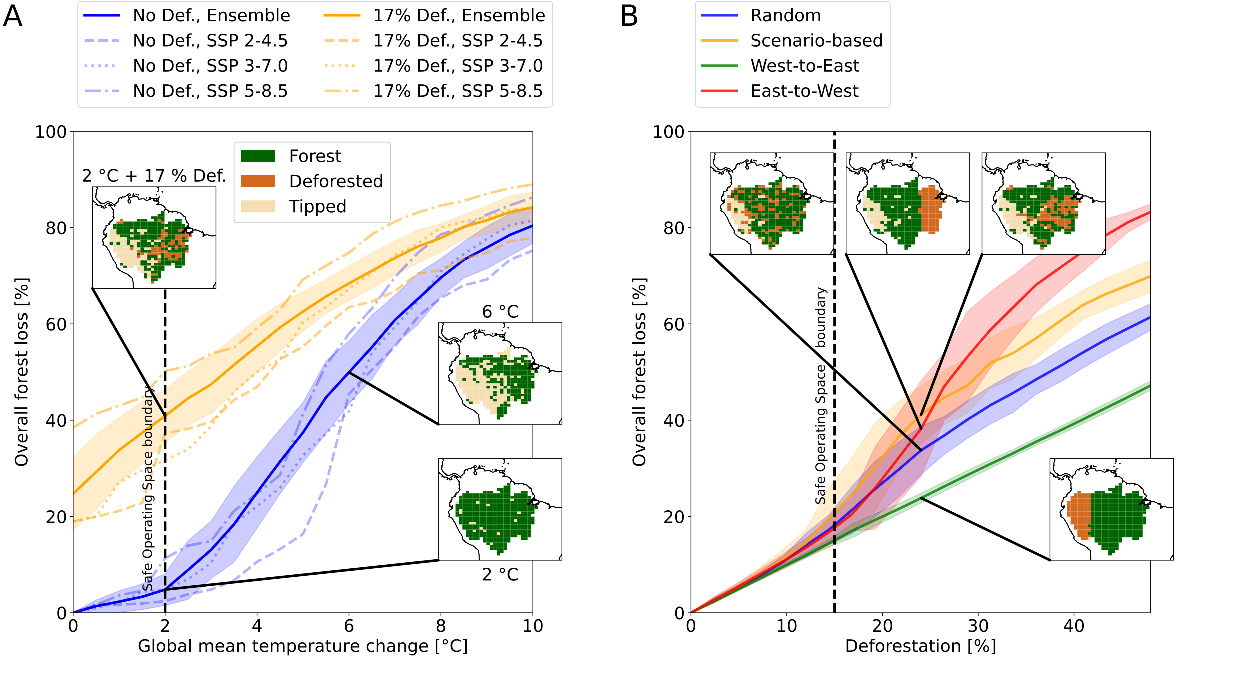}
  \caption{\textbf{Individual response of Amazon rainforest to the drivers.} (\textbf{A}) Overall forest loss with respect to global warming without deforestation (blue) and with 17~\% deforestation (orange). The straight line indicates the mean of the ensemble. The shaded area shows the 95~\% confidence interval determined from the results of 16 ensemble members. The dashed lines show the response according to the parameters determined from each individual scenario. Without deforestation, the effects of global warming become detrimental at about 2~$^\circ$C as indicated by the dashed line. With deforestation, the detrimental effects already unfold especially in the west of the Amazon basin (see inset). (\textbf{B}) Overall forest loss with respect to deforestation without global warming. Nonlinear effects differ between spatial deforestation pattern (see inset maps) but start at about 15~\% of deforestation as indicated by the dashed line.}
  \label{fig:climate-change}
\end{figure}

We observe a nonlinear response with respect to both drivers. For climate change (Fig. \ref{fig:climate-change}A), the overall forest loss starts to increase at global mean temperature changes above $2~^\circ$C without deforestation. For deforestation only (Fig. \ref{fig:climate-change}B), we observe nonlinear increases of overall forest loss starting at around $15~\%$ deforestation for all except the west-to-east deforestation pattern (the different deforestation patterns are discussed below). Both values are compatible with the minima of tipping ranges from the literature (deforestation is even slightly lower than assumed by \cite{flores_critical_2024}). Therefore, we consider these values as our first estimation of the safe boundaries of the Amazon rainforest.\par
However, we can see from Fig. \ref{fig:climate-change}A that at 17~$\%$ deforestation the overall forest loss increases with further climate change already starting at 0~$^\circ$C. This confirms that the estimation of individual boundaries is, in general, not sufficient to quantify the safe operating space of the Amazon rainforest as these can give a false impression of safety. Significant forest tipping mainly occurs in the western part of the Amazon basin because these are the regions that receive most of the recycled moisture and therefore might experience exceptionally strong forest losses in response to pressures on the system as a whole.

\subsection*{Combined safe operating space of the Amazon rainforest}
\begin{figure}
  \centering
  \includegraphics[width=0.8\textwidth]{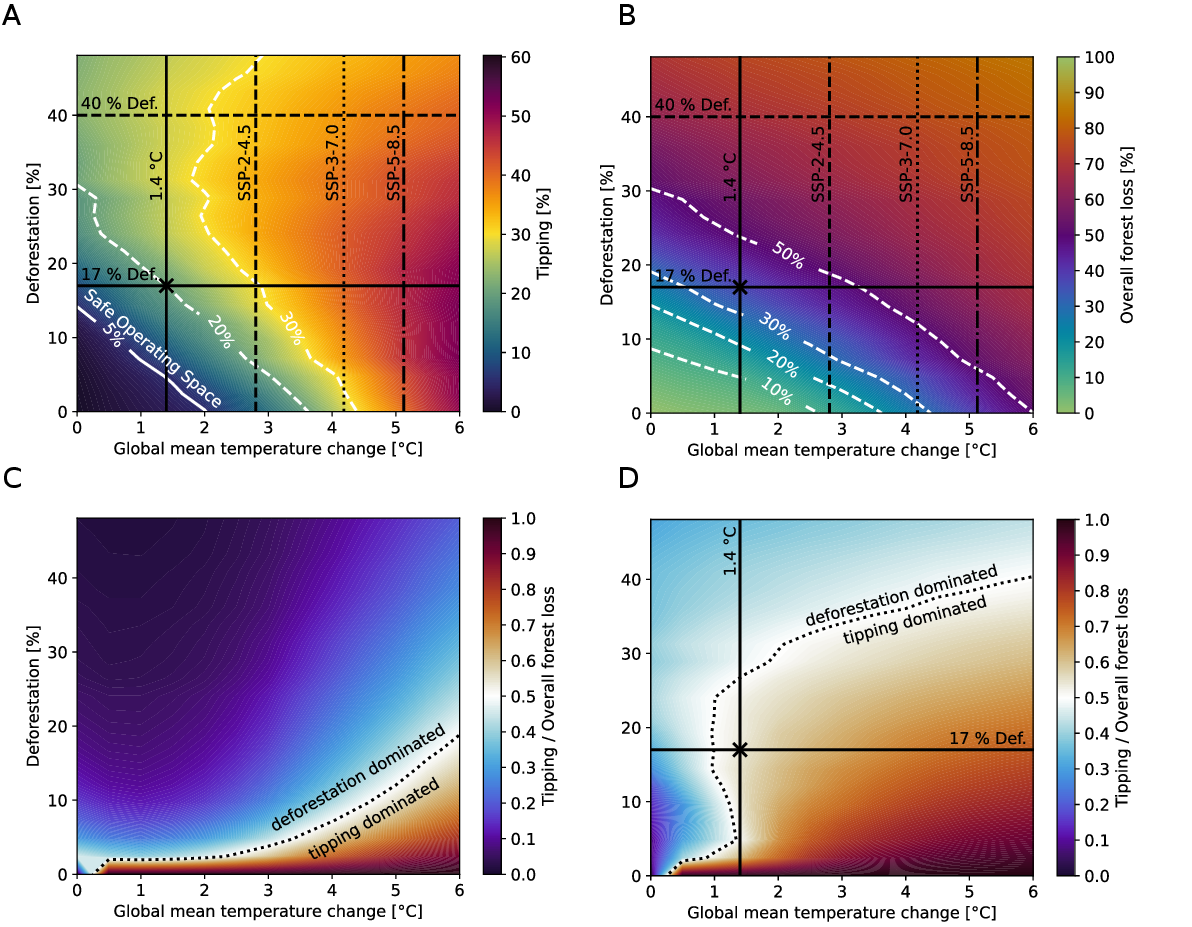}
  \caption{\textbf{Safe Operating Space of the Amazon rainforest for the ensemble run with the scenario-based deforestation pattern.} (\textbf{A}) Tipping (without deforestation) with respect to deforestation and global mean temperature change. The white line shows the boundary of the safe operating space and a higher tipping risk zone. The vertical lines indicate different global warming levels for present day and the emission scenarios at 2100. The horizontal lines indicate the current deforestation level and the expected deforestation level at 2050 according to the scenario-based projections \cite{soares-filho_modelling_2006}. The present-day values of global warming and deforestation are indicated with a cross. (\textbf{B}) Overall forest loss with respect to deforestation and global mean temperature change. At present-day values, 17~\% deforestation and 20~\% tipping add up to an overall forest loss of 37~\% (\textbf{C} and \textbf{D}) Fraction of tipping over forest loss at different levels of global mean temperature change and deforestation showing the relative abundance of tipping without moisture-recycling (\textbf{C}) and with moisture-recycling (\textbf{D}). A value of $0.5$ (dotted line) indicates that deforestation and climate change contribute equally to the overall forest loss. Without moisture recycling, the space is separated into a deforestation-dominated and a tipping-dominated region indicated by blue areas and red areas, respectively. With moisture recycling, the line of equal contribution bends backwards and the tipping-dominated region significantly expands.}
  \label{fig:sos}
\end{figure}

For a combined assessment, we quantify the full operating space (Fig. \ref{fig:sos}A). We find that our individual safe boundary assessments of $2~^\circ\mathrm{C}$ and 15~\% deforestation above correspond to tipping of 5~\% of the forest area, which we thus find as safe boundary of the safe operating space. The critical temperature of the safe boundary decreases sharply with deforestation, indicating a strong interaction between climate change and deforestation.\par
For climate change of 3-4~$^\circ\mathrm{C}$, corresponding to the best estimate of a tipping point \cite{armstrongmckay_exceeding_2022}, we observe 20~\% tipping in the absence of deforestation. Present-day climate change does not, in isolation, transgress the boundary of the safe operating space. For the combination of present-day global warming ($1.4~^\circ\mathrm{C}$) and deforestation (17~\%), however, the safe operating space is transgressed mainly due to deforestation. In fact, the present-day value is located on the 20~\% tipping isoline which puts more than 30~\% of the Amazon rainforest at risk (see Fig. \ref{fig:sos}B). Note however, that we do not observe system-wide tipping here. In addition, we define a high risk zone at tipping of $T>30~\%$. The three scenarios all reach or even enter the high risk zone for current and projected deforestation values. In contrast, deforestation projections alone do not transgress into the high risk zone due to a saturation of the detrimental effects of deforestation at $D\approx 30~\%$.\par
We then look at the fraction of tipping over overall forest loss with and without moisture recycling to compare the relative effects of climate change and deforestation in different parts of the operating space. Without moisture recycling most of the condition levels are deforestation dominated (Fig. \ref{fig:sos}C). With moisture recycling (Fig. \ref{fig:sos}D) tipping-dominated condition levels increase significantly and expand into global-warming levels below $2~^\circ\mathrm{C}$ for deforestation levels of $5-25~\%$. Due to this, present-day climate change and deforestation is located within the tipping-dominated condition levels, indicating a particularly high vulnerability of the Amazon rainforest to climate change at current levels of global warming and deforestation (see Fig. \ref{fig:sos}D).\par
Our model evaluation (fig. \ref{supfig:evaluation}) experiments show that average precipitation is more relevant for exceeding global-warming induced tipping points, while seasonal water deficit is more relevant for tipping events that originate from cascading moisture recycling reductions initiated by deforestation.
In summary, the results show that the current state of the Amazon rainforest is not only out of its safe operating space but could also currently be in a particularly dangerous region due to the present accumulated deforestation and global warming.
\subsection*{Detrimental pathways of deforestation}
We further investigate how different spatial patterns of deforestation influence the safe operating space of the Amazon rainforest. Maps for the different deforestation patterns at 17~\% deforestation can be seen in Fig. \ref{fig:climate-change}B and the corresponding safe operating space assessments in fig. \ref{supfig:sos}. Overall, we find that east-to-west and west-to-east deforestation patterns correspond to a worst- and best-case, and the random deforestation pattern corresponds to an intermediate case. Thus, the results for these patterns, albeit highly artificial, are an ideal gauge for the scenario-based deforestation pattern we focus on.\par
In Fig. \ref{fig:climate-change}B, the response of the Amazon rainforest to deforestation only (i.e without global warming) is shown for all deforestation patterns. Except for the west-to-east scenario, all scenarios show a nonlinear response of additional tipping to a varying degree. This can be attributed to the moisture recycling that mainly acts from east to west. Interestingly, while for large deforestation (more than 25~\%) overall forest loss is highest for the east-to-west pattern, for small deforestation (less than 25~\%) the overall forest loss for the scenario-based pattern is highest. An explanation could be that, for the scenario-based pattern, forest loss occurs closer to the vulnerable regions in the west. This initial advantage for small deforestation is outweighed by a large and consistent accumulated deforestation in the east. This result suggests that the actual deforestation pattern might be close to the worst possible pattern for present and possible future deforestation levels. \par 
Comparing the safe operating spaces for the different deforestation patterns (see fig. \ref{supfig:sos}) shows that almost no synergistic effects between global warming and deforestation occur for the west-to-east pattern, i.e. the impacts of climate change and deforestation are almost completely independent. For the other scenarios, moderate (random) to high (east-to-west) synergy between global warming and deforestation can be observed. This synergy can be explained with the direction of the moisture recycling where tipping of cells in the east of the Amazon basin leads to cascading destabilization of cells in the west of the Amazon basin. The fraction of tipping over overall forest loss confirms this (see fig. \ref{supfig:fraction}): For the patterns with moderate to high synergy (random, scenario-based and east-to-west) the tipping-dominated region is expanded as discussed above. In contrast, for the west-to-east deforestation pattern, the curve looks similar to the one without moisture recycling in Fig. \ref{fig:sos}C. It should be kept in mind that while this suggests that no nonlinear response to deforestation might be expected for west-to-east deforestation, significant deforestation will nonetheless have detrimental impacts on the local ecosystems of the Amazon rainforest.\par
\begin{figure}
\centering
\includegraphics[width=1\textwidth]{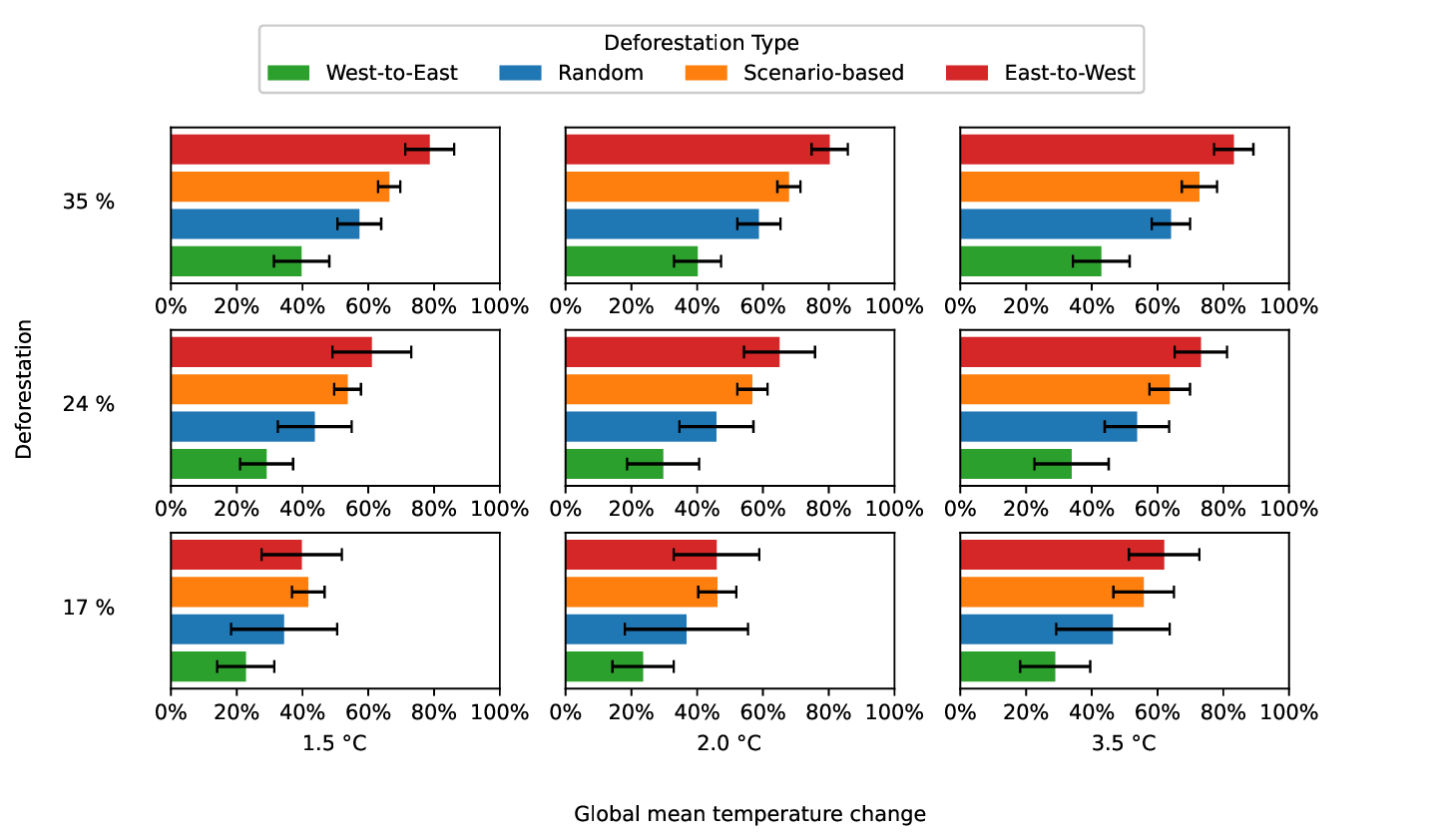}
\caption{\textbf{Comparison of the effect of the deforestation patterns for different values of global warming and deforestation.} Each bar shows the mean overall forest loss of the ensemble for the corresponding deforestation pattern indicated by the color. The error bars indicate the standard deviation across the ensemble. In almost all cases, the forest loss in the scenario-based pattern is between the forest loss in the intermediate random pattern and the forest loss in the worst-case east-to-west pattern. Only for lower (present-day) values of global warming and deforestation, the overall forest loss is even higher than the forest loss from the east-to-west deforestation pattern.}
\label{fig:comparison}
\end{figure}
Comparing the overall forest loss (combination of deforestation and tipping) among the different patterns across the operating space (Fig. \ref{fig:comparison}) shows that the scenario-based deforestation pattern consistently lies between the random (intermediate case) and the east-to-west (worst case) for almost all values of climate change and deforestation except for low to intermediate climate-change and low deforestation. This is similar to what was observed for deforestation only. Overall, our results confirm the hypothesis that the scenario-based deforestation leads to particularly high resilience losses and that the western parts of the Amazon are highly vulnerable to forest dieback in the eastern parts of the Amazon basin due to moisture recycling.

\subsection*{Discussion}
We quantified the safe operating space of the Amazon rainforest under climate change and deforestation \cite{bultan_amazon_2025} using a reduced-complexity model and explicitly consider atmospheric moisture recycling. We found a transition from mostly forest covered to mostly degraded state starting at around $2~^\circ$C in line with recent evidence on Amazon tipping \cite{melnikova_amazon_2025}. Synergistic effects between global warming and deforestation are found due to atmospheric moisture recycling. These effects are comparatively strong for present-day accumulated deforestation of 17~\% along a scenario-based deforestation pattern, suggesting that current deforestation pathways are particularly dangerous for the resilience of the Amazon rainforest. We furthermore found that, due to deforestation, the effects of climate change could be detrimental well before estimated climate change tipping points are crossed, that is, at present-day climate change and deforestation levels.\par
However, one of the assumptions of this study, the existence and location of local-scale tipping points, remains uncertain, particularly on different spatial scales \cite{wuyts_amazonian_2017, zwaan_widespread_2024}. While a parameterization based on past climate variability is reasonable and based on biological mechanisms, combining an empirically grounded approach similar to \cite{nian_potential_2023} with explicit consideration of moisture recycling, as in our study, might be a significant improvement for further studies. For this, a combination of remote-sensing data and derived early-warning signals might be used to parameterize local tipping points. Early-warning signs are often also used to quantify resilience and show decreasing resilience across the Amazon basin \cite{boulton_pronounced_2022}. However, the interpretation of early-warning signs in observational data also raises a number of conceptual and statistical challenges \cite{rietkerk_ambiguity_2025}.\par
Furthermore, we employed a reduced-complexity model where complexity is mostly introduced in terms of the data-based accounting for moisture recycling. On the one hand, this simplicity comes with the ability to easily parameterize the model to data, run ensembles and quantify uncertainties. On the other hand, the model does not directly include physical or biological mechanisms. An important line of future work would therefore be the investigation of Earth System Models and the abundance of tipping points in them using standardized approaches, such as TIPMIP \cite{winkelmann_tipping_2025}. However, these models would need to simulate vegetation dynamically and properly include the important feedback mechanisms of fire and moisture recycling. Additionally, our analysis is also relevant for other systems where moisture recycling plays a role, such as Congo, South East Asian rainforests or boreal forests. Our approach could be extended to also assess these regions.\par
Overall, the results we present show a worrying synergy between climate change and deforestation. The current state of the Amazon rainforest appears highly vulnerable, which leads to the conclusion that we may have already left the safe operating of the Amazon rainforest. Further climate change or deforestation can thus have drastic and potentially irreversible consequences for the rainforest ecosystem and its functions. This underlines the vital importance of international efforts to stop climate change and deforestation and return to a safe operating space.


\clearpage 

%
\bibliography{amazon_sos} 
\bibliographystyle{amazon_sos}

%
%
%
%
%
%


\section*{Acknowledgments}
\paragraph*{Funding:} 
This is ClimTip contribution \#156; the ClimTip project has received funding from the European Union's Horizon Europe research and innovation programme under grant agreement No. 101137601: Funded by the European Union. Views and opinions expressed are however those of the author(s) only and do not necessarily reflect those of the European Union or the European Climate, Infrastructure and Environment Executive Agency (CINEA). Neither the European Union nor the granting authority can be held responsible for them. A.~S. would like to acknowledge funding from the ERC-Synergy project RESILIENCE, proposal no. 101071417. We gratefully acknowledge the European Regional Development Fund, the German Federal Ministry of Education and Research, and the Land Brandenburg for supporting this project by providing resources on the high-performance computer system at the Potsdam Institute for Climate Impact Research.

\paragraph*{Author contributions:}

N.W. conceived the study. N.W. and J.K. designed the study. J.K. wrote the code and analysed the results. All authors discussed and interpreted results. J.K. wrote the manuscript with input from all authors.
\paragraph*{Competing interests:}
There are no competing interests to declare
\paragraph*{Data and materials availability:}
Code and data needed to evaluate and reproduce the results are available at https://doi.org/10.5281/zenodo.19913489. The moisture recycling data are available at https://doi.org/10.5281/zenodo.10650579. The deforestation data are available at https://doi.org/10.3334/ORNLDAAC/1153. No new materials have been created for this work.


\subsection*{Supplementary materials}
Materials and Methods\\
Figs. S1 to S7\\
References \textit{(50-\arabic{enumiv})}\\ 


\newpage


\renewcommand{\thefigure}{S\arabic{figure}}
\renewcommand{\thetable}{S\arabic{table}}
\renewcommand{\theequation}{S\arabic{equation}}
\renewcommand{\thepage}{S\arabic{page}}
\setcounter{figure}{0}
\setcounter{table}{0}
\setcounter{equation}{0}
\setcounter{page}{1} 


\begin{center}
\section*{Supplementary Materials for\\ \scititle}

	Jonathan.~Krönke,$^{\ast}$
	Arie.~Staal,
	Jonathan.~F.~Donges,
    Johan.~Rockström,
    Nico.~Wunderling$^{\ast}$\\
    \small$^\ast$Corresponding authors. Emails: jonathan.kroenke@pik-potsdam.de, wunderling@c3s.uni-frankfurt.de\\
\end{center}

\subsubsection*{This PDF file includes:}
Materials and Methods\\
Figures S1 to S7\\

\newpage


\subsection*{Materials and Methods}
\subsubsection*{Environmental data}
Similar to \cite{wunderling_pinpointing_2025} we used 10-year averaged (to exclude the influence of single years) precipitation and evaporation output of NorESM2 \cite{seland_overview_2020} for the period of 1950-2014 and for three scenarios (SSP2-4.5,  SSP3-7.0 and SSP5-8.5) to estimate the effect of global warming on the environmental conditions in a region. The data have a resolution of $1.25^\circ\times0.9375^\circ$. From that, we computed the mean annual precipitation and the maximum cumulative water deficit \cite{aragao_spatial_2007} for each cell and hydrological year (October to September) with
\begin{equation}
    p = \frac{1}{12}\sum_{n=\mathrm{Oct.}}^{\mathrm{Sep.}} p_n
\end{equation}
and
\begin{align}
    s &= \max(d_n, d_{n+1}, ..., d_{n+11}) \\
    d_n &= d_{n-1} + e_n - p_n \\
    \min(d_n) &= 0
\end{align}
with precipitation $p$, evaporation $e$, cumulative water deficit $d$ and maximum cumulative water deficit $s$.
\subsubsection*{Global warming}
We used the median global temperature change as simulated in MAGICC7 [see fig. 4.40a of \cite{ipcc_climate_2021}] to map the global mean temperature for the corresponding scenarios to the years and conducted a linear regression of the mean annual precipitation and maximum cumulative water deficit with respect to global mean temperature change, thus obtaining a relation of the environmental conditions under climate change for each cell $i$
\begin{align}
    p_i(T) = p_i^0 + \gamma_i T, \\
    s_i(T) = s_i^0 + \delta_i T.
    \label{eq:relation}
\end{align}
Maps of the mean annual precipitation and maximum cumulative water deficit at a global mean temperature change of $T=0^\circ$C and the slopes $\gamma_i$ and $\delta_i$ can be seen in Fig. \ref{supfig:offset} and Fig. \ref{supfig:slope}, respectively.
\subsubsection*{Moisture recycling networks}
We used UTrack \cite{tuinenburg_tracking_2020}, a Lagrangian moisture-tracking algorithm applied to the NorESM2 data \cite{seland_overview_2020}. For each mm of evaporation in a grid cell, 100 parcels of moisture are released at random places above that cell. The position is then updated according to the three-dimensional wind speeds and directions with a time step of four hours. If precipitation occurs at a time step within a grid cell, the moisture content of parcels is updated and counted as precipitation until only 1~\% of the original moisture in the parcel remains or after 30 days. This results in a matrix $R$ of precipitation values $r_{ij}$ in cell $i$ that originate from cell $j$ for each month. A more thorough explanation of the algorithm is provided in \cite{staal_global_2025}.\par
\begin{figure} 
	\centering
	\includegraphics[width=0.9\textwidth]{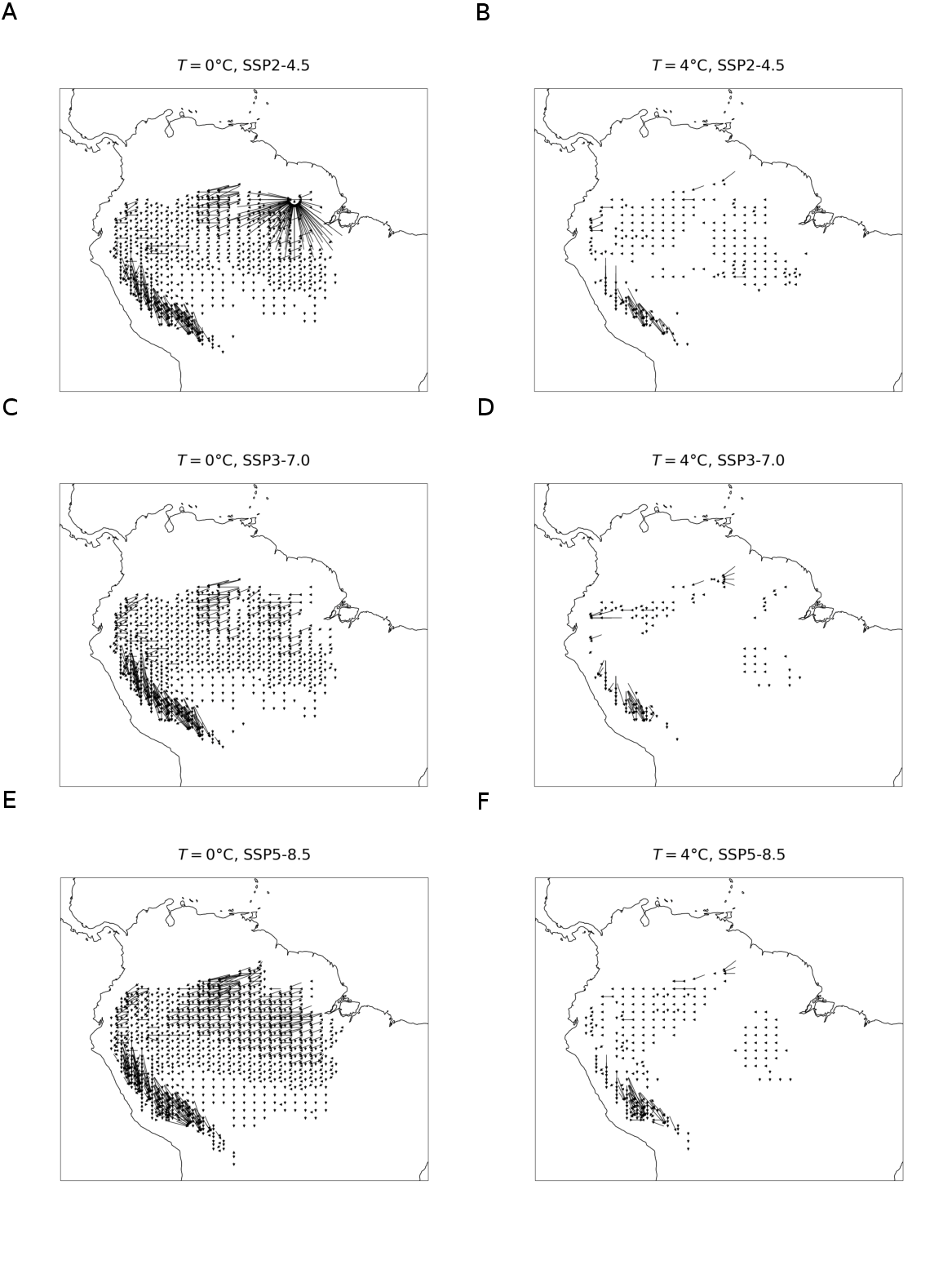} 

	\caption{moisture recycling networks for the different scenarios. For better visualization, only links with $\Delta p_{ij}>15$~mm/yr or $\Delta s_{ij}>2$~mm/yr are shown. With increasing global mean temperature the moisture recycling decreases due to decreasing precipitation.}
	\label{supfig:networks}
\end{figure}
\begin{figure} 
	\centering
	\includegraphics[width=1\textwidth]{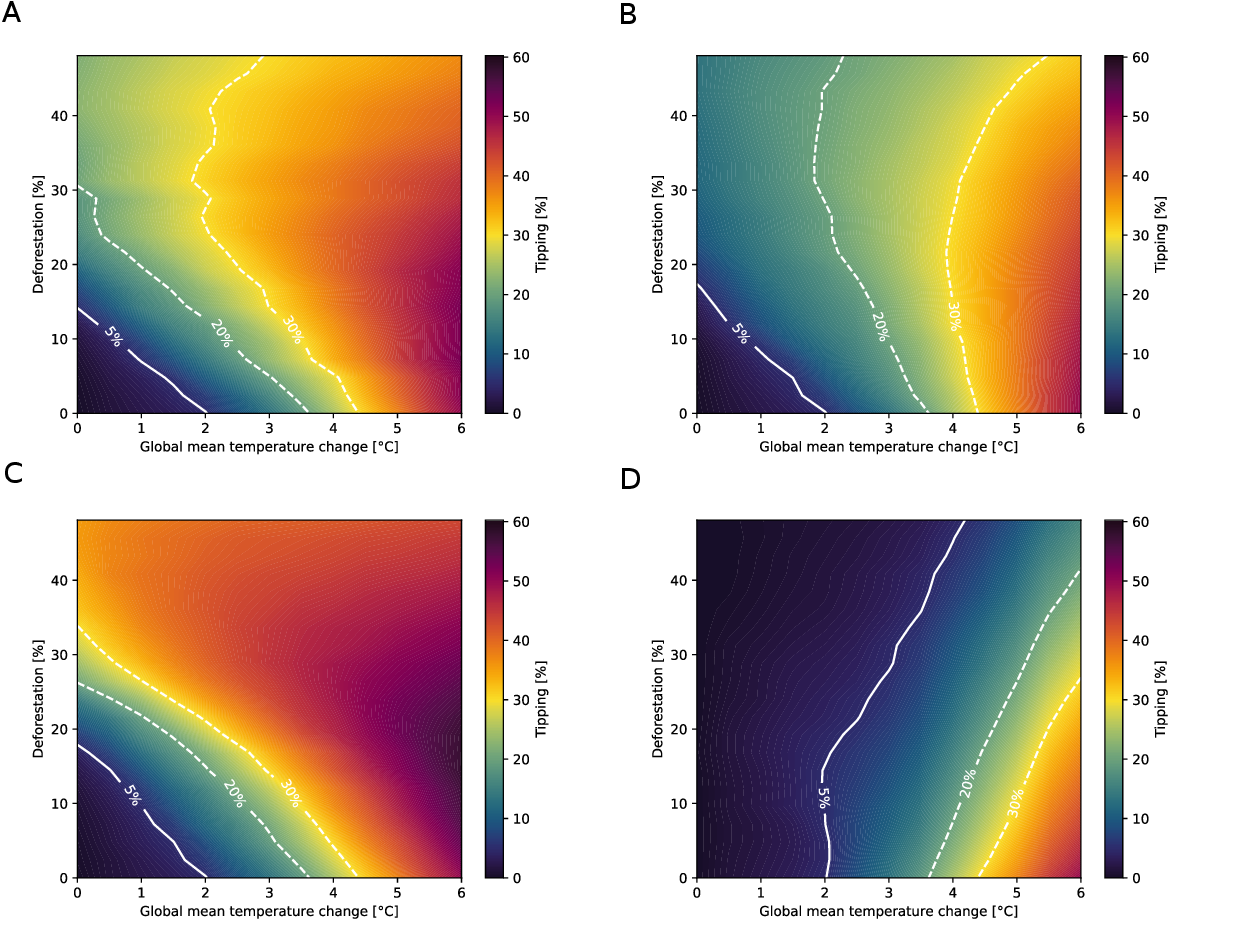} 

	\caption{\textbf{Same as Figure~\ref{fig:sos}A, but for all deforestation patterns.}
		(\textbf{A}) scenario-based, (\textbf{B}) random, (\textbf{C}) east-to-west, (\textbf{D}) west-to-east. For the west-to-east deforestation pattern, no synergistic effects between climate and deforestation occur. For the other patterns, the synergistic effects are pronounced.}
	\label{supfig:sos}
\end{figure}
\begin{figure} 
	\centering
	\includegraphics[width=1\textwidth]{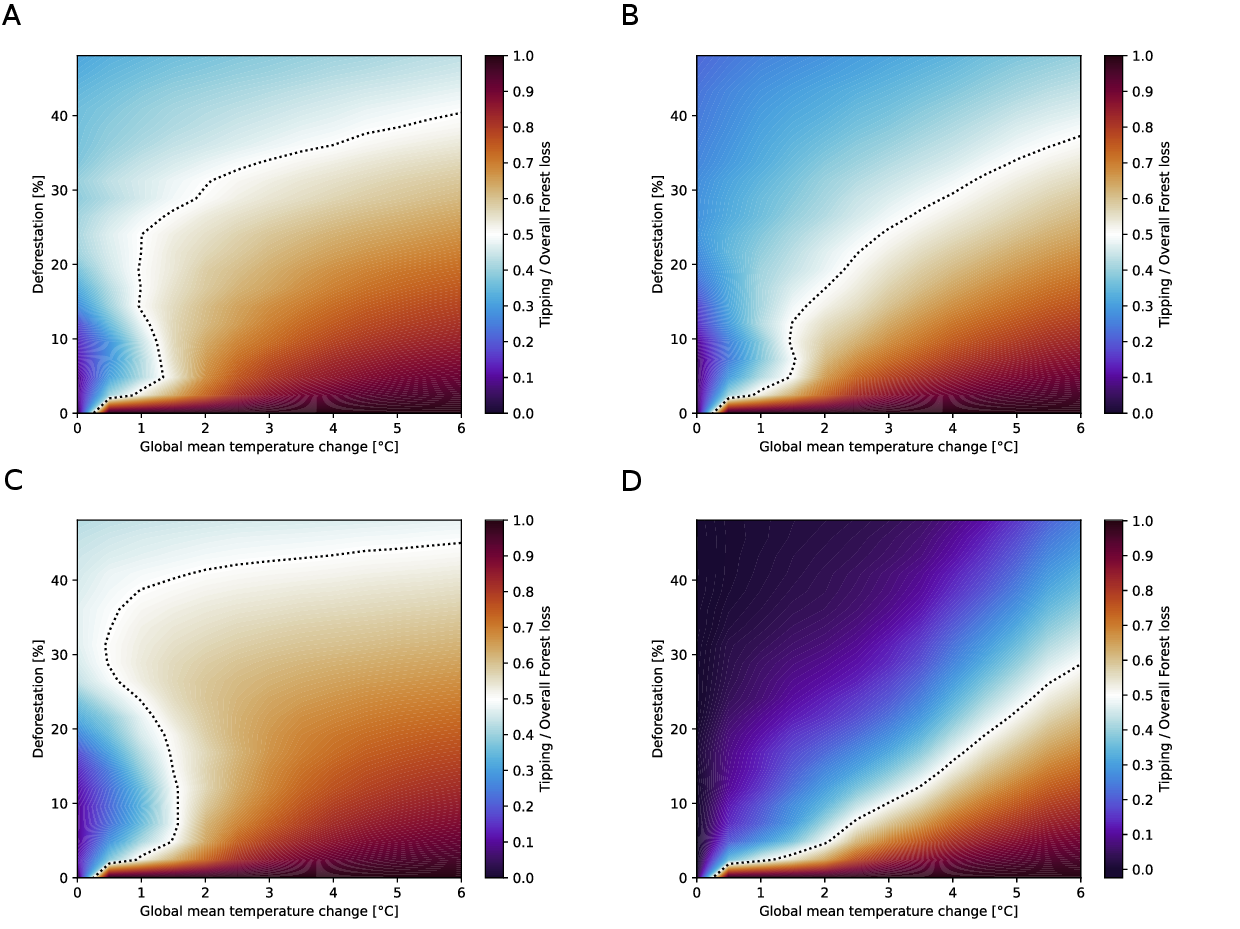} 

	\caption{\textbf{Same as Figure~\ref{fig:sos}D, but for all deforestation patterns.}
		(\textbf{A}) scenario-based, (\textbf{B}) random, (\textbf{C}) east-to-west, (\textbf{D}) west-to-east.}
	\label{supfig:fraction}
\end{figure}
\begin{figure} 
	\centering
	\includegraphics[width=1\textwidth]{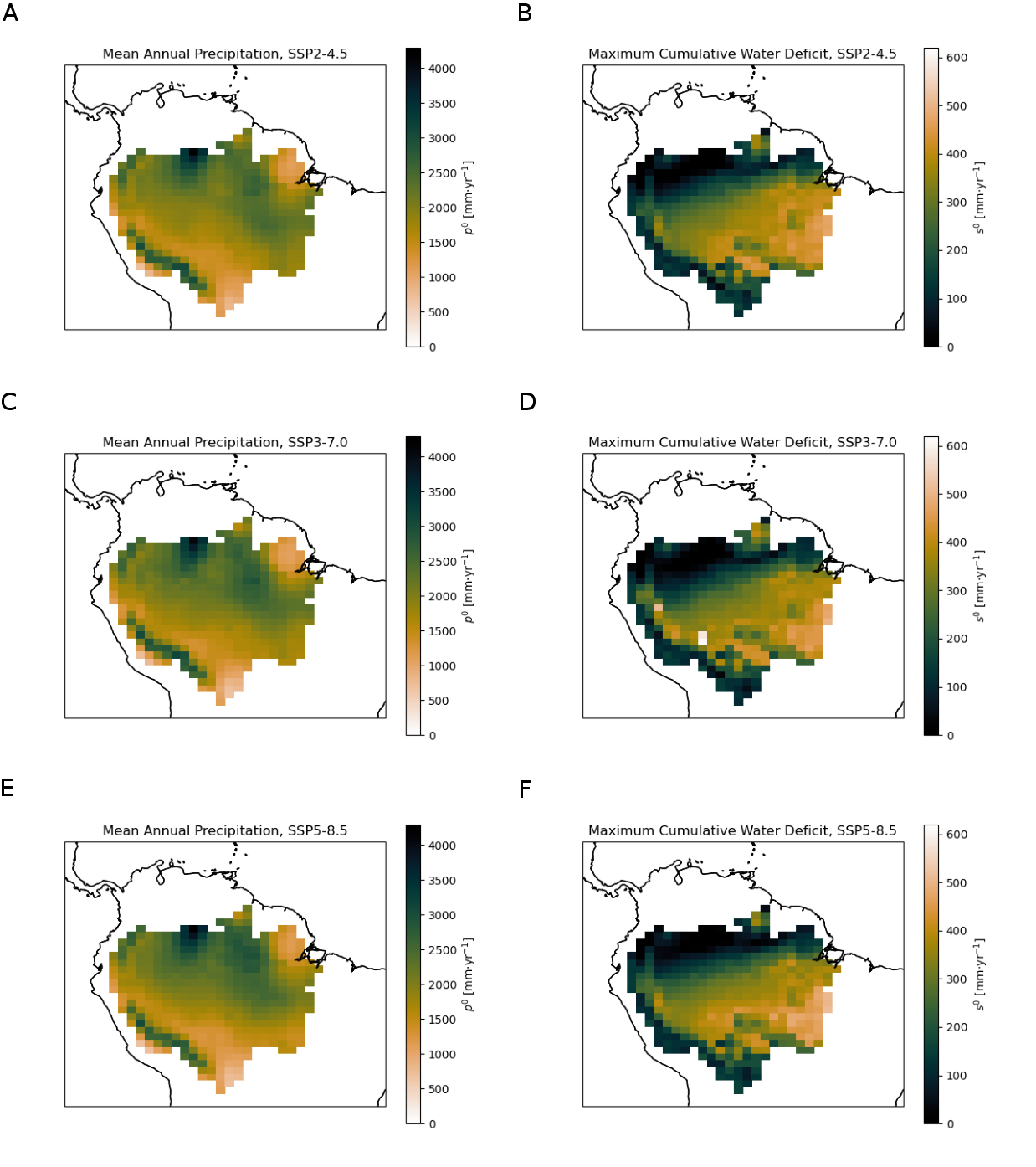} 

	\caption{Mean annual precipitation and maximum cumulative water deficit for $\Delta T =0^\circ$C for the different scenarios.}
	\label{supfig:offset}
\end{figure}
\begin{figure} 
	\centering
	\includegraphics[width=1\textwidth]{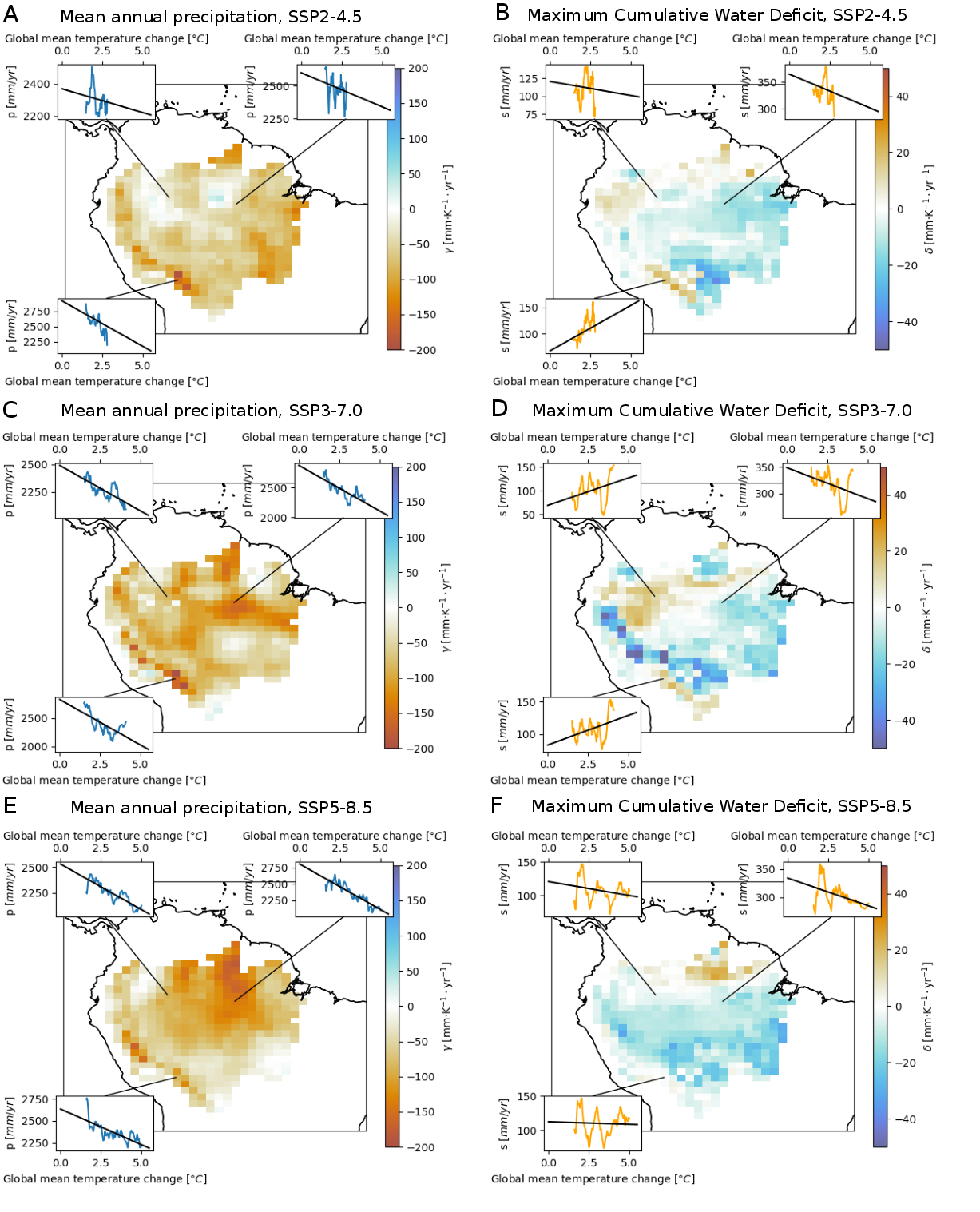} 

	\caption{Response of mean annual precipitation and maximum cumulative water deficit to global warming for the different scenarios.}
	\label{supfig:slope}
\end{figure}

For this study, we then subtracted these values from the monthly precipitation values and calculated the hypothetical mean annual precipitation and maximum cumulative water deficit in cell $j$ removing the link from each cell $i$. The difference between the hypothetical and original values results in the matrices $S$ of maximum cumulative water deficit differences $\Delta s_{ij}$ and $P$ of mean annual precipitation differences $\Delta p_{ij}$. For each link between two cells, we performed the same linear regression as for the local environmental conditions to obtain relations $\Delta s_{ij}(T)$ and $\Delta p_{ij}(T)$ for the three scenarios. Examples of moisture recycling networks only showing links above a threshold of either $\Delta p_{ij} > 15$~mm or $\Delta s_{ij}> 2$~mm for the different scenarios at $T=0~^\circ$C and $T=4~^\circ$C can be seen in Fig. \ref{supfig:networks}.
\subsubsection*{Adaptation}\label{sec:adaptation}
We assume that plants are adapted to the local environmental conditions. To account for this, we parametrize local tipping points with respect to past climate variability. For this, we use results of the historical simulation run of NorESM2 from which the used climate change scenarios are branched. The standard deviation for the economic drivers mean annual precipitation and maximum cumulative water deficit between 1950 and 2014 are calculated for each cell individually. The estimated critical values then read
\begin{align}
    p^\mathrm{crit}_i&=\bar{p_i}-\beta\sigma_i^\mathrm{p}, \\
    s^\mathrm{crit}_i&=\bar{s_i}+\beta\sigma_i^\mathrm{s},
\end{align}
with the historical mean values $\bar{s}$ and $\bar{p}$, the historical standard deviation $\sigma^\mathrm{s}$ and $\sigma^\mathrm{p}$ and a scaling factor $\beta$ that is set to $\beta=1.0$ during our study. The critical values are then used in Eqs. \ref{eq:local} and \ref{eq:interaction}, effectively resulting in a tipping point when the change in the corresponding driver reaches one standard deviation.
\subsubsection*{Dynamical system}
We use an established approach based on a conceptual tipping model \cite{wunderling_recurrent_2022, wunderling_modelling_2021, kronke_dynamics_2020}. The Amazon rainforest is modeled as interacting dynamical system following the equation
\begin{equation}
    \frac{dx_i}{dt}=-x_i^3+x_i+\sqrt{\frac{4}{27}}\cdot r_i(x_1,...,x_N),
    \label{eq:methods:cusp}
\end{equation}
where $i=1,...,N$ are the grid cells and $x_i$ the state of the cell $i$. This equation is the normal form of a fold-bifurcation with two stable states, forest ($x_i < 0$) and alternative state ($x_i > 0$), e.g. savanna or dry forest, assuming the system is in equilibrium. Depending on the environmental conditions $r_i$, the forest state is stable $r_i < 1$ or unstable $r_i > 1$ in which case the local forest cell will tip into the stable alternative state.\par
The environmental conditions $r_i$ are determined by
\begin{equation}
    r_i=f_\mathrm{local}(p_i,s_i)+\sum_{j,j \neq i}^N f_\mathrm{interaction}(\Delta p_{ij}, \Delta s_{ij}, x_j).
    \label{eq:environmental_conditions}
\end{equation}
The first term accounts for the local environmental conditions and is determined by the mean annual precipitation $p_i$ and the maximum cumulative water deficit $s_i$ in that cell. It reads
\begin{equation}
    f_\mathrm{local}(p_i, s_i)=\frac{s_i(T) - s_i(0)}{s_i^\mathrm{crit} - s_i(0)} + \frac{p_i(T) - p_i(0)}{p_i^\mathrm{crit} - p_i(0)}.
    \label{eq:local}
\end{equation}
Neglecting the second term in Eq. \ref{eq:environmental_conditions}, the cell is at its tipping point when $f(p_i,s_i)=1$, i.e. if either $s_i$ increases by one standard deviation of the historical record, $p_i$ decreases by one standard deviation of the historical record or a combination of both. This is a slight variation from the base study for the sake of model simplicity. Furthermore, if the first term in Eq. \ref{eq:local} is smaller than zero, it is set to zero to exclude unrealistic resilience increases due to an decrease of maximum cumulative water deficit.\par
The second term in Eq. \ref{eq:environmental_conditions} corresponds to the effect of tipping in one cell $j$ on the environmental conditions in another cell $i$. It is calculated with
\begin{equation}
    f_\mathrm{interaction}(\Delta p_{ij}, \Delta s_{ij}. x_j)=\left(\frac{\Delta s_{ij}}{s_i^\mathrm{crit} - s_i(0)}-\frac{\Delta p_{ij}}{p_i^\mathrm{crit} - p_i(0)}\right)\frac{x_j+1}{2},
    \label{eq:interaction}
\end{equation}
For the forest state $x_j \approx -1$ the effect on the environmental conditions of cell $i$ are approximately zero. For the tipped state ($x_j \approx 1$) we calculate the environmental conditions in cell $i$ subtracting the difference in mean annual precipitation ($\Delta p_{ij}$) and maximum cumulative water deficit ($\Delta s_{ij}$) that results from the missing moisture recycling from cell $j$. The coupling strength is then chosen, such that tipping of cell $j$ leads to such a change in the environmental conditions of cell $i$. For computational reasons we discarded all values where both $\Delta p_{ij} < 1\,\mathrm{mm}$ and $\Delta s_{ij} < 0.1\,\mathrm{mm}$ were true in the simulations resulting in a complex network of moisture recycling with a link density of around 10~\%.
\subsubsection*{Deforestation scenarios}
Deforestation is not uniformly distributed across the Amazon basin. Therefore, we used four different spatial deforestation patterns (see Fig. \ref{fig:climate-change}B). The primary pattern was a scenario-based pattern from simulations that considers road and infrastructure expansion as well as deforestation \cite{soares-filho_modelling_2006}. We used the data for 2050 and sorted our grid cells by the amount of deforestation in them. A certain percentage of deforestation then equates to the state of the corresponding cells in Eq. \ref{eq:methods:cusp} being in the alternative state ($x_i = 1$) as initial condition to our simulations.\par
In addition to the scenario-based pattern, we employed an east-to-west, a west-to-east and a random deforestation pattern. For the west-to-east deforestation pattern, all cells in the eastern columns are considered to be in the alternative state first until a certain percentage of deforestation is reached. If in one column not all cells are considered to be deforested, the deforested cells with the same latitude were chosen randomly. For the east-to-west deforestation pattern, the same method was used choosing the western columns first.
\subsubsection*{Ensemble and simulation procedure}
\label{sec:ensemble}
To simultaneously account for randomness in the deforestation patterns and uncertainties in the response to global warming we used an ensemble approach. For each global mean temperature change and deforestation, we simulated an initial value problem for Eq. \ref{eq:methods:cusp} with an ensemble of $N=16$ members. For each member, the initial condition was chosen according to the deforestation pattern. The response of mean annual precipitation and maximum cumulative water deficit (Eq. \ref{eq:relation}) for each cell was calculated for all three scenarios. From this, a normal distribution was calculated and the parameters for each cell were drawn from this. Overall this results in a total number of 56448 simulations.
\subsubsection*{Robustness checks and model evaluation}
\begin{figure} 
	\centering
	\includegraphics[width=1\textwidth]{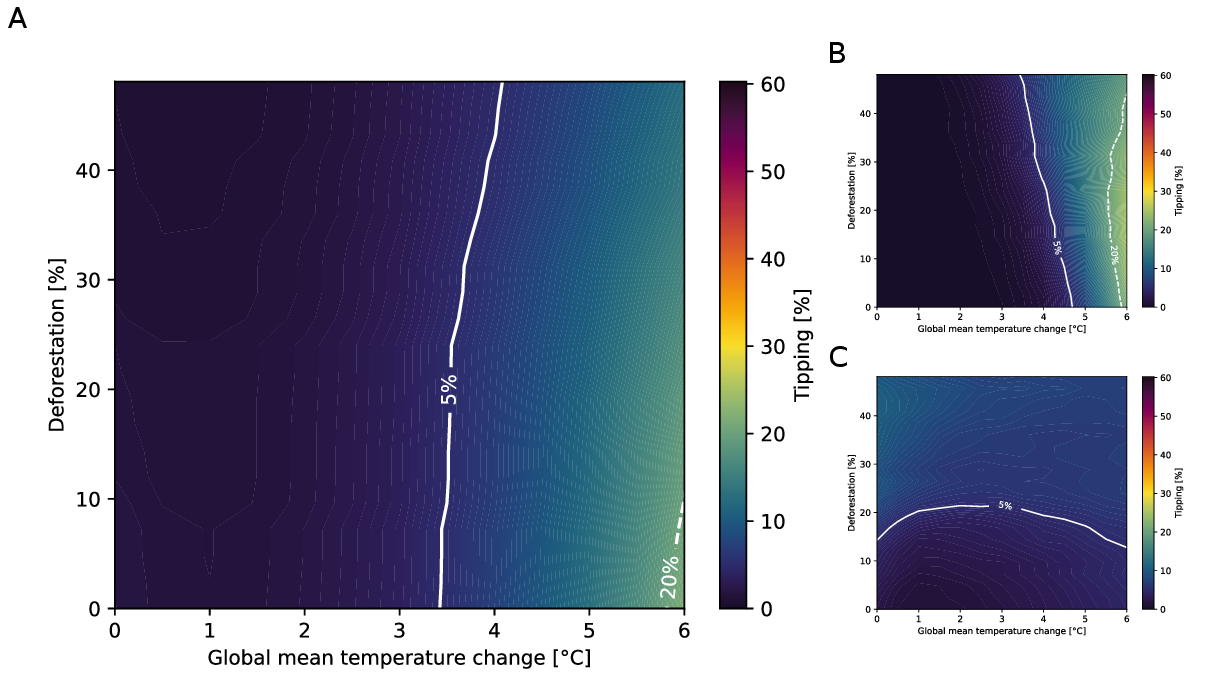} 

	\caption{\textbf{Same as Figure~\ref{fig:sos}A, but leaving out different mechanisms.}
		(\textbf{A}) Leaving out moisture recycling, (\textbf{B}) only considering mean annual precipitation, and (\textbf{C}) only considering maximum cumulative water deficit}
	\label{supfig:evaluation}
\end{figure}

\begin{figure} 
	\centering
	\includegraphics[width=1\textwidth]{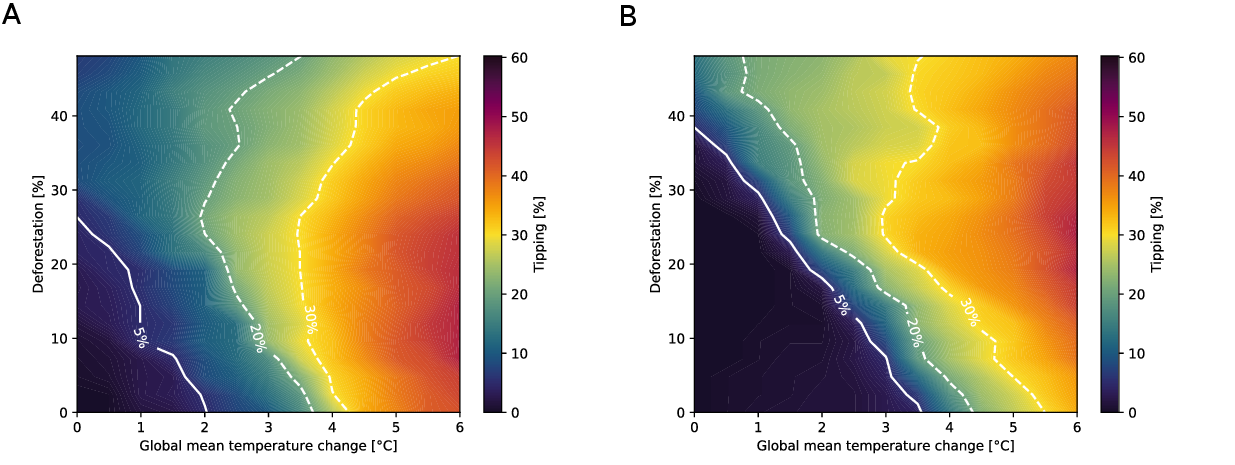} 

	\caption{\textbf{Same as Figure~\ref{fig:sos}A, but for the robustness checks.}
		(\textbf{A}) Using fixed thresholds in addition to adaptive thresholds, and (\textbf{B}) using a fixed evaporation of 100~mm in each grid cell.}
	\label{supfig:robustness}
\end{figure}

Additionally, we evaluated the model by leaving out different effects, namely leaving out moisture recycling and considering only the effects of mean annual precipitation and maximum cumulative water deficit (Fig. \ref{supfig:evaluation}). The results show again that moisture recycling strongly exacerbates the effects of global warming. However, while mean annual precipitation is more relevant for exceeding global-warming induced tipping points, maximum cumulative water deficit is especially relevant for tipping events that come from cascading moisture recycling reductions initiated by deforestation.  
To evaluate the robustness of our results we conducted the following checks (Fig. \ref{supfig:robustness}):
First, we used fixed thresholds of $p^\mathrm{crit}=1000~\mathrm{mm}$ and $s^\mathrm{crit}=450~\mathrm{mm}$ (according to \cite{flores_critical_2024}) in addition to the adaptive thresholds. Second, we use a fixed evaporation of 100~mm instead of the NorESM2 data in each grid cell to calculate the maximum cumulative water deficit. Both checks show that our results are qualitatively robust.\par



\end{document}